\documentclass[onecolumn,aps,showpacs,prb,preprint,superscriptaddress]{revtex4}
\usepackage{tipa}
\usepackage{bbding}
\usepackage{txfonts}
\usepackage{amssymb}
\usepackage{graphicx}
\usepackage{CJK}

\begin{document}


\title{The thermodynamic and kinetic properties of hydrogen dimers on graphene}

\author{Liang Feng Huang}\affiliation{Key
Laboratory of Materials Physics, Institute of Solid State Physics,
Chinese Academy of Sciences, Hefei 230031, China}
\author{Mei Yan Ni}\affiliation{Key Laboratory of
Materials Physics, Institute of Solid State Physics, Chinese Academy
of Sciences, Hefei 230031, China}\affiliation{School of Electronic
Science and Applied Physics, Hefei University of Technology, Hefei
230009, China}
\author{Yong Gang Li}
\author{Wang Huai Zhou}
\author{Xiao Hong Zheng}
\author{Ling Ju Guo}
\author{Zhi Zeng}\thanks{Email: zzeng@theory.issp.ac.cn}\affiliation{Key Laboratory of Materials Physics, Institute of
Solid State Physics, Chinese Academy of Sciences, Hefei 230031,
China}

\begin{abstract}
The thermodynamic and kinetic properties of hydrogen adatoms on
graphene are important to the materials and devices based on
hydrogenated graphene. Hydrogen dimers on graphene with coverages
varying from 0.040 to 0.111 ML (1.0 ML $=
3.8\times10^{15}$cm$^{-2}$) were considered in this report. The
thermodynamic and kinetic properties of H, D and T dimers were
studied by ab initio simulations. The vibrational zero-point energy
corrections were found to be not negligible in kinetics, varying
from 0.038 (0.028, 0.017) to 0.257 (0.187, 0.157) eV for H (D, T)
dimers. The isotope effect exhibits as that the kinetic mobility of
a hydrogen dimer decreases with increasing the hydrogen mass. The
simulated thermal desorption spectra with the heating rate $\alpha =
1.0$ K/s were quite close to experimental measurements. The effect
of the interaction between hydrogen dimers on their thermodynamic
and kinetic properties were analyzed in detail.
\end{abstract}


\maketitle

\section{INTRODUCTION}
\par The thermodynamic and kinetic properties of hydrogen (isotopes) on graphene can help
us understand the interaction between hydrogen and graphitic
materials in outer space\cite{coey,hornekaer96,hornekaer97,ferro368}
and fusion devices.\cite{ferro368,morris} Such properties also
determine the realizability of the graphene(graphite)-based hydrogen
storage\cite{anthony} and the graphene-based electronic
devices.\cite{shytov,ryu,elias,sofo75,balog131,bostwick103,guisinger9,luo3,lebegue79}
And the hydrogenated graphene (graphite) has been one focus of
scientific researches in recent years.

\par Thermal desorption (TD) spectroscopy, which is also called temperature-programmed desorption
(TPD) spectroscopy, has been used in experiments to study the
kinetic properties of hydrogen on the graphite
surface.\cite{hornekaer96,zecho42,zecho117,andree425} T. Zecho et
al.\cite{zecho42,zecho117} found that the saturation coverages
($\Theta_{sat}$) of H and D on graphite surface are about 0.3 and
0.4 ML, respectively. The TD spectra of H and D (heating rate
$\alpha = 1.0$ K/s) both exhibit two-peak shapes when the coverage
$\Theta \leq 0.3\times\Theta_{sat}$, and the increase in $\Theta$
just results in some minor modification of the two-peak shapes. L.
Horkek{\ae}r et al. found that there are mainly two kinds of
hydrogen dimers responsible for the two-peak shapes of the TD
spectra at low hydrogen coverage, which are the ortho-dimer and
para-dimer. In the two-peak shaped TD spectra, the two desorption
peaks are at 445 (490) K and 560 (580) K for H (D) adatoms,
respectively,\cite{zecho42,zecho117} where the positions of the
desorption peaks increase with the hydrogen mass. F. Dumont et al.
have simulated the TD spectra of hydrogen dimers on graphene by
using the kinetic Monte Carlo method,\cite{dumont77} and reproduced
the two-peak shaped TD spectra. And E. Gavardi et al. also have
simulated the two-peak shaped TD spectra by the same method but
taken hydrogen trimers and tetramers into
consideration.\cite{gavardi477} These experimental and theoretical
results indicate that the hydrogen ortho-dimer and para-dimer should
be still the most important dimers responsible for the near-two-peak
shapes of the TD spectra at high coverages. This is consistent with
that the configurations of the ortho-dimer and para-dimer always
tend to be preserved in hydrogen trimers and
tetramers.\cite{gavardi477,roman78}

\par However, the kinetic properties of hydrogen dimers still need
more accurate simulations, especially the vibrations should be taken
into consideration. Because the mass of hydrogen isotopes are so
small that the vibrational zero-point energy corrections may not be
negligible in kinetics. The thermodynamic and kinetic properties of
hydrogen dimers are expected to be sensitively dependent on their
microscopic structures, on which systematic and accurate
investigations are still lacking.

\par In this report, the thermodynamic and kinetic properties of H, D and T dimers on
graphene are simulated with a composite method consisting of density
functional theory (DFT),\cite{kohn140} density functional
perturbation theory (DFPT),\cite{baroni73} harmonic transition state
theory (hTST),\cite{eyring3,vineyard3,hanggi62,pollak15,toyoura78}
and reaction rate theory.\cite{hanggi62} The effect of the
interaction between hydrogen dimers on the thermodynamic and kinetic
properties of hydrogen dimers on graphene are studied. The isotope
effects in the kinetic properties are observed due to the inclusion
of the vibrations in the simulations. The simulated TD spectra of H
and D dimers with the heating rate $\alpha = 1.0$ K/s are very close
to experimental measurements, and the TD spectra of T (radioactive)
dimers are also predicted.

\section{METHODOLOGY}
\par The adsorption energy of a hydrogen dimer is defined to be the energy
difference before and after the adsorption of two hydrogen atoms on
graphene, which is expressed as
\begin{eqnarray}{\label{adsorption}}
E_{ads}=2E^H+E^{GL}-E^{GL+2H}
\end{eqnarray}
where $E^H$, $E^{GL}$ and $E^{GL+2H}$ are the total electronic
energies of an isolated hydrogen atom, an isolated graphene layer
and a graphene layer with a chemisorbed hydrogen dimer on it,
respectively.

\par The frequency ($v$) of an over-barrier jump from one local
minimum state (initial state) to another local minimum state (final
state) can be calculated using the quantum-mechanically modified
harmonic transition state theory (hTST), which is expressed
as\cite{eyring3,toyoura78}
\begin{eqnarray}{\label{qm_hTST}}
v & = & v^*_{qm}\exp{(-\frac{E_{ac}}{k_BT})}\nonumber\\
  & = & \frac{k_BT}{h}\frac{\prod\limits_{i=1}^{3N}[1-\exp{(-\frac{\hbar\omega_i^I}{k_BT})}]}{\prod\limits_{i=1}^{3N-1}[1-\exp{(-\frac{\hbar\omega_i^S}{k_BT})}]}\exp{(-\frac{E_{ac}}{k_BT})}
\end{eqnarray}
where $v^*_{qm}$ is the quantum-modified exponential prefactor;
$E_{ac}$ is the activation energy; N is the number of atoms;
$\omega_i^I$ and $\omega_i^S$ are the frequencies of the
{\itshape{i}}th vibrational mode in the initial and saddle-point
states in the reaction path, respectively. The numbers $3N$ and
$3N-1$ denote that the initial state has $3N$ real vibrational modes
and the saddle-point state has $3N-1$ real modes with 1 imaginary
mode excluded in the calculation. The activation energy here is
defined to be the vibrational zero-point energy corrected potential
barrier, which is expressed as
\begin{eqnarray}{\label{activation}}
E_{ac} & = & {\Delta}E_{p}+\frac{1}{2}\sum_{i=1}^{3N-1}\hbar{\omega}_i^S-\frac{1}{2}\sum_{i=1}^{3N}\hbar{\omega}_i^I \nonumber\\
       & = & {\Delta}E_{p}-{\Delta}F_{vib}(0)
\end{eqnarray}
where ${\Delta}E_p$ is the potential barrier in the reaction path;
${\Delta}F_{vib}(0)$ is the vibrational zero-point energy
correction. When the temperature approaches to be infinite, the
classical limit of the prefactor is expressed as
\begin{eqnarray}{\label{cl_hTST}}
v^*_{cl}=\frac{1}{2\pi}\frac{\prod\limits_{i=1}^{3N}\omega_i^I}{\prod\limits_{i=1}^{3N-1}\omega_i^S}
\end{eqnarray}
This classical-limit form is also the Vineyard's
form,\cite{vineyard3} where all of the vibrational modes are assumed
to be completely thermo-activated.

\par
\begin{figure}[!h!t!bp]
\scalebox{0.5}[0.5]{\includegraphics{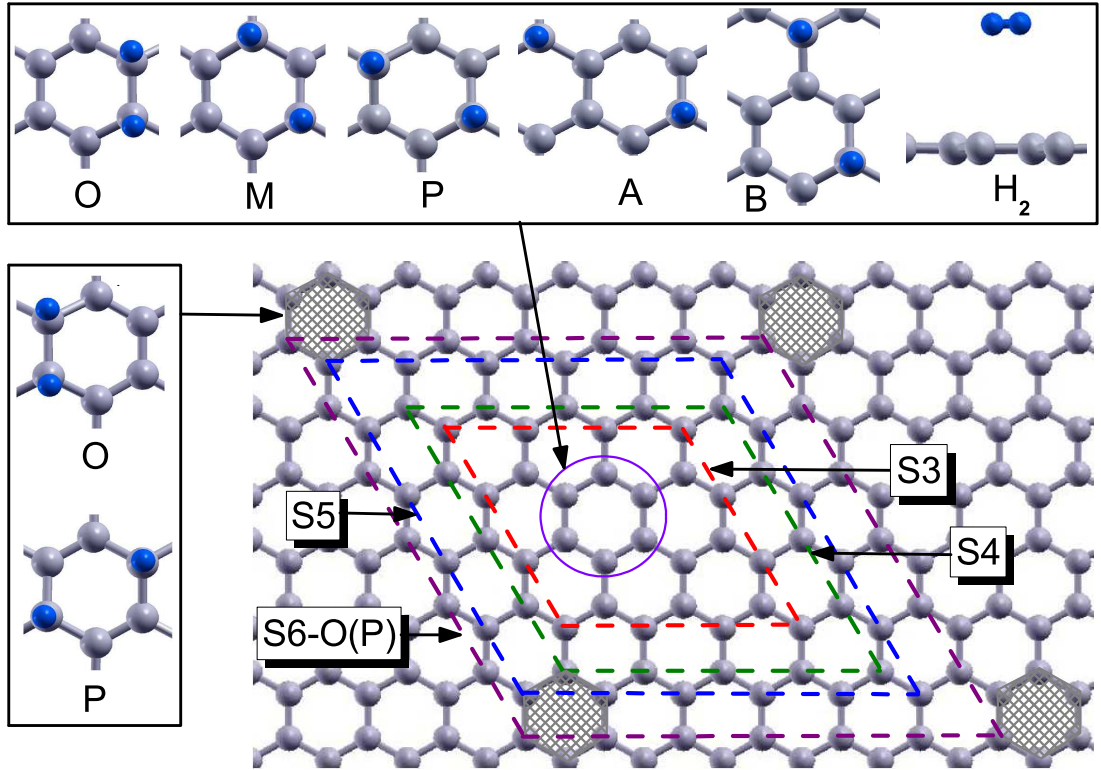}}
\caption{\label{structures}(Color online) The structures of the
periodic graphene supercells with the sizes of $3\times3$ (S3, 18 C
atoms), $4\times4$ (S4, 32 C atoms), $5\times5$ (S5, 50 C atoms) and
$6\times6$ (S6, 72 C atoms) times of the graphene unit cell. A
hydrogen dimer is allowed to be chemisorbed at the center of each
supercell (marked by a purple circle). There are 5 configurations of
hydrogen dimers on graphene considered here, which are ortho-dimer
(O), meta-dimer (M), para-dimer (P), A-dimer (A) and B-dimer (B), as
well as the desorbed hydrogen molecule (H$_2$). And the S6 graphene
supercell is decorated with a para-dimer (S6-P) or an ortho-dimer
(S6-O) at the corner (shaded in gray) of the supercell. The gray
spheres are carbon atoms, and the blue smaller spheres are hydrogen
atoms.}
\end{figure}

 In this report, a mono-layer graphene is used as a structural
model, which is also a safe model for graphite surface because the
weak Van de Waals interaction between neighboring graphene layers
has negligible effect on the chemisorption properties of
hydrogen.\cite{hornekaer96,hornekaer97,lfhuang1,sljivancanin131} The
vacuum between two neighboring layers is 10 \AA{}. And $\Theta$s of
0.111, 0.063 and 0.040 ML (1.0 ML $= 3.8\times10^{15}$cm$^{-2}$) are
computationally achieved by chemisorbing a hydrogen dimer at the
center of periodic graphene supercells with sizes of $3\times3$
(S3), $4\times4$ (S4), and $5\times5$ (S5) times of the graphene
unit cell, respectively (Fig. \ref{structures}). Besides,  a
para-dimer decorated $6\times6$ supercell (S6-P) and an ortho-dimer
decorated $6\times6$ supercell (S6-O), as shown in Fig.
\ref{structures}, are designed to allow a hydrogen dimer chemisorbed
at the supercell center ($\Theta = 0.056$ ML). The choice of the
para- and ortho-dimer for decoration is due to that the two dimers
were found to be the most stable hydrogen dimers on
graphene.\cite{sljivancanin131} In experimental observations,
hydrogen adatoms (including dimers) tend to cluster on graphite
surface,\cite{hornekaer96,hornekaer97} making some spots covered by
hydrogen while some not. This un-uniform distribution of hydrogen
indicates that the $\Theta$ in simulation should correspond to the
effective hydrogen coverage of the hydrogen-covering spots. There
are 5 configurations of hydrogen dimers on graphene considered here,
which are ortho-dimer (O), meta-dimer (M), para-dimer (P), A-dimer
(A) and B-dimer (B) as shown in Fig. \ref{structures}, as well as
the desorbed hydrogen molecule (H$_2$). Other more extended hydrogen
dimers\cite{sljivancanin131} are not considered.

\par The structures and potential barriers are calculated using DFT and
the vibrational frequencies using DFPT. The DFT and DFPT
calculations are carried out using the Quantum Espresso code
package,\cite{giannozzi21} in which the ultrasoft\cite{vanderbilt41}
pseudopotentials with the BLYP\cite{becke38,lee37}
exchange-correlation functional are used. The energy cutoffs for the
wave function and charge density are 35 and 350 Ry, respectively.
Uniform k-point grids are chosen to be $6\times6$, $5\times5$,
$4\times4$ and $3\times3$ for the S3, S4, S5 and S6-P(O) cases,
respectively. The Methfessel-Paxton smearing
technique\cite{methfessel40} with an energy width of 0.03 Ry is
employed to speed the convergence of the numerical calculations. The
electronic density of states (DOS) for the chemisorption states in
the S5 case is calculated with a $6\times6$ k-point grid and the
tetrahedron interpolation scheme.\cite{blochl49} The reaction paths
are described by the minimum energy paths (MEPs) between two local
minimum states, which are calculated using the climbing-image nudged
elastic band method.\cite{henkelman113} For the calculation of the
vibrational frequencies, only the $\Gamma$ point at the Brillouin
zone center is chosen.

\section{RESULTS AND DISCUSSION}
\subsection{DFT and hTST calculations}
\par The calculated adsorption energies for the P-dimer ($E_{ads}^P$) and
O-dimer ($E_{ads}^O$) on the S3, S4, S5, S6-P and S6-O graphene
supercells are listed in Tab. \ref{E_ads}. In the S3, S4 and S5
cases, $E_{ads}$ generally increases with $\Theta$, with the only
exception that the $E^O_{ads}$ in the S3 case is a little (9 meV)
less than that in the S4 case. This indicates the attractive
interaction between the neighboring dimers, which is consistent with
the clustering of hydrogen adatoms on graphite surface observed in
experiments.\cite{hornekaer96,hornekaer97} This is because one
chemisorbed dimer can cause the carbon lattice around wrinkled, and
partly destroy the $sp^2$ orbital hybridization of the carbon atoms
nearby, which can increase the affinity of those carbon atoms to
bond with another hydrogen dimer. However, the $E_{ads}^P$ in the
S6-P(O) case is larger than those both in the S4 and S5 cases, and
the $E_{ads}^O$ in the S6-P(O) case is larger than those in all the
S3, S4 and S5 cases, although the $\Theta$ of the S6-P(O) case is
lower than those of the S3 and S4 cases. This is due to the effect
of the periodic boundary conditions (PBC) we take for the DFT
calculations here. In the S3, S4 and S5 cases, two interacting
dimers are the periodic images of each other. While in the S6-P and
S6-O cases, a dimer can interact with the decoration dimer at the
corner of the same supercell. Thus, in the S3, S4 and S5 cases, the
PBC will partly suppress the wrinkling of the carbon lattice caused
by the adsorption of a dimer, while in the S6-P and S6-O cases, a
dimer feels more wrinkling caused by the decoration dimer in the
same supercell. Thus, this PBC effect on the thermodynamic stability
of hydrogen dimers on graphene also approves that the wrinkling of
the carbon lattice will result in the attractive interaction between
dimers on graphene.

\par The calculated potential barriers (${\Delta}E_p$) for various
transitions of the hydrogen dimers on the S3, S4, S5, S6-P and S6-O
supercells are listed in Tab. \ref{potential_barrier}. It can be
seen that ${\Delta}E_p$ is dependent on both $\Theta$ and the
structural configuration, the same as $E_{ads}^P$ and $E_{ads}^O$
described in the previous paragraph. In all cases, the optimal
reaction path for an O-dimer to be desorbed is O-M-P-H$_2$, and that
for the P-dimer is P-H$_2$, which is consistent with Hornek\ae{r}'s
results.\cite{hornekaer96,sljivancanin131} Other transitions out of
the O-M-P-H$_2$ reaction path should overcome higher potential
barriers, e.g. ${\Delta}E_p$(M-B) is 0.189 eV larger than
${\Delta}E_p$(M-P) in the S6-P case (much larger in other four
cases). This magnitude of difference will result in that the
possibility of the M-P transition is about 3 orders larger than that
of the M-B transition at 300 K (estimated by Equ. \ref{qm_hTST}).
Thus, the M-B transition could be safely neglected when compared
with the M-P transition. In addition, according to the calculations
by Hornek\ae{r} and
\v{S}ljivan\v{c}anin,\cite{hornekaer96,hornekaer97,sljivancanin131}
the evaporation of hydrogen monomers from dimers also needs overcome
much higher potential barriers. Therefore, only the transitions in
the optimal path O-M-P-H$_2$ are considered in the simulation of the
TD spectra in the following. For an isolated dimer on graphene,
there may be more than one paths that are equivalent for one kind of
transition, e.g. there are 4 equivalent paths for the O-M
transition. The number of equivalent paths for a transition is
defined as the degeneracy of the reaction path ($g_{path}$). The
values of the $g_{path}$s for the transitions of an isolated dimer
on graphene are listed in Tab. \ref{potential_barrier}, where the
$g_{path}$s for the adsorbing transitions (H$_2$-M, H$_2$-P and
H$_2$-O) are absent, because they are hardly and unnecessarily
defined in the simulations here. However, when $\Theta$ is high
enough, the interaction between dimers will tend to make two
different paths of the same kind transition have different
$\Delta{E}_p$s, namely the $g_{path}$ is reduced. Among the
transitions in the optimal path, the P-H$_2$ and O-M transitions are
the most important ones (the rate-limiting steps) for the TD
spectra, which will be shown in the following. The
${\Delta}E_p$(O-M) and ${\Delta}E_p$(P-H$_2$) generally increase
with $\Theta$ in the S3, S4 and S5 cases, which indicates that it is
harder for hydrogen dimers to be desorbed from graphene at higher
$\Theta$s. This is the same as the effect of $\Theta$ on the
thermodynamic stability of hydrogen dimers on graphene, as discussed
in the previous paragraph. However, the ${\Delta}E_p$(O-M) in the
S6-O case and the ${\Delta}E_p$(P-H$_2$) in the S6-O(P) case are
larger than those in the S3, S4 and S5 cases. This is because the
PBC effect tend to make the dimers on a S6-P(O) supercell more
stable than those on the S3, S4 and S5 supercells. Thus, it costs
more energy for a dimer in the S6-P(O) case to escape from one state
to transit to another state.

\par
\begin{figure}[!h!t!bp]
\scalebox{0.3}[0.3]{\includegraphics{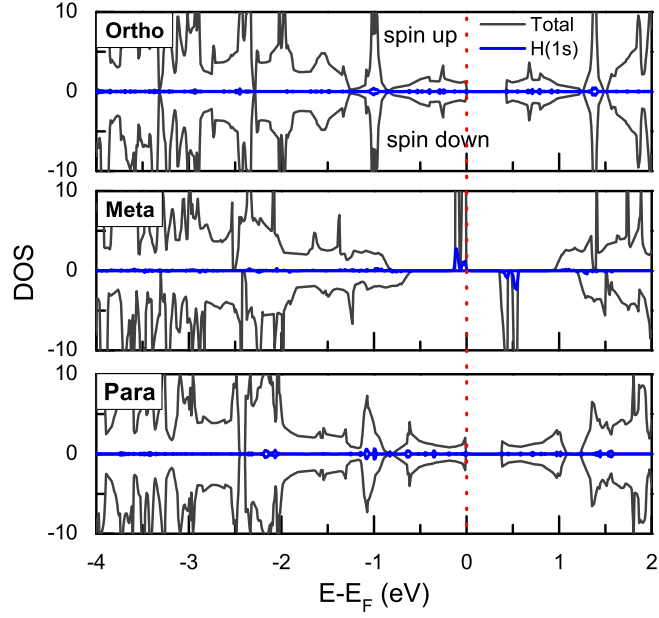}}
\caption{\label{DOS}(Color online) The DOS spectra of the S5
graphene supercell with an O-dimer, M-dimer and P-dimer. The
vertical line at the Fermi level guides the eyes.}
\end{figure}

 The adsorption of a hydrogen dimer onto two C atoms in graphene will change
these C atoms from $sp^2$ to $sp^3$ hybridized. Seen from the DOS
spectra for the systems of S5+O-dimer, S5+M-dimer and S5+P-dimer
(Fig. \ref{DOS}), the hybridization between hydrogen dimers and
graphene makes graphene be semiconducting. The band gap are 0.43,
0.36 and 0.38 eV for S5+O-dimer, S5+M-dimer and S5+P-dimer,
respectively. However, the DOS spectra for S5+M-dimer are different
from those for S5+O-dimer and S5+P-dimer. Only in the DOS spectra of
S5+M-dimer, there are four narrow peaks around the Fermi level, two
spin-up (occupied) and two spin-down (unoccupied). And S5+M-dimer is
magnetic with a spin moment of 2.0 $\mu_B$, while the other two are
nonmagnetic. The two C atoms bonded with M-dimer are equivalent in
graphene, a bipartite system, while those C atoms bonded with
O(P)-dimer are inequivalent. The difference in magnetism is due to
this difference in the hydrogen-bonded sites, which can be
understood from the Lieb's theorem for the bipartite
system.\cite{lieb62} The adsorption of a M-dimer leaves two
unsaturated $p_z$ orbitals, which then form two quasilocal states
around the M-dimer, with each state holding one electron and a spin
moment of 1.0 $\mu_B$. This is the same as the adsorption of the
hydrogen monomer on graphene.\cite{lfhuang1,casolo130,casolo81} The
contributions of the H($1s$) orbitals to these quasilocal states are
very small. However, the two unsaturated $p_z$ orbitals in
S5+O-dimer or S5+P-dimer bond with each other, thus, there are no
quasilocal states and magnetic moment in these two systems. This
bonding between these two $p_z$ orbitals makes the O-dimer and
P-dimer be more stable than the M-dimer (Tab.
\ref{potential_barrier}). Although the transition of a hydrogen
dimer changes the magnetism of the hydrogenated graphene, the system
is dominated by nonmagnetic states. Because the M-dimer is
metastable (seen from the $\Delta{E}_p$s in Tab.
\ref{potential_barrier}), and its lifetime at 300 K is less than 1.0
s, as simulated below. The two hydrogen atoms of a dimer prefer to
adsorb on two inequivalent C atoms in graphene. Likewise, from the
energetic calculations for more extended
dimers\cite{sljivancanin131} and trimers\cite{roman78} on graphene,
a configuration is also much more stable when two
nearest-neighboring hydrogen atoms are bonded with inequivalent C
atoms in graphene.

\par The calculated ${\Delta}F_{vib}(0)$s for the transitions of
H, D and T dimers in the optimal path O-M-P-H$_2$ are listed in Tab.
\ref{ZPE_corrections}, where the results for the S3 and S5 cases are
compared. The differences between these two groups are very small,
especially  for the ${\Delta}F_{vib}(0)$s of the two rate-limiting
transitions, O-M and P-H$_2$ (within 6 meV). This implies that the
vibrational properties of the hydrogen dimers on graphene are mainly
determined by the localized vibrational modes, which is the same as
those of the hydrogen monomer on graphene.\cite{lfhuang1} It also
can be seen that ${\Delta}F_{vib}(0)$ decreases with increasing the
hydrogen mass. This isotope effect is also the same as that of the
hydrogen monomer on graphene in Ref. \onlinecite{lfhuang1}, where
the relationship between the isotope effects in phonon spectra and
${\Delta}F_{vib}(0)$ was analyzed in detail. The
${\Delta}F_{vib}(0)$s for the diffusing transitions (O-M, M-O, M-P,
P-M) are less than those for the desorbing transition (P-H$_2$),
which is also the same as that of the hydrogen monomer on
graphene.\cite{lfhuang1} The values of the ${\Delta}F_{vib}(0)$s for
H (D, T) dimer vary from 0.038 (0.028, 0.017) to 0.257 (0.187,
0.157) eV, which are large enough that can not be omitted in
kinetics.

\par
\begin{figure}[!h!t!bp] 
\scalebox{0.18}[0.18]{\includegraphics{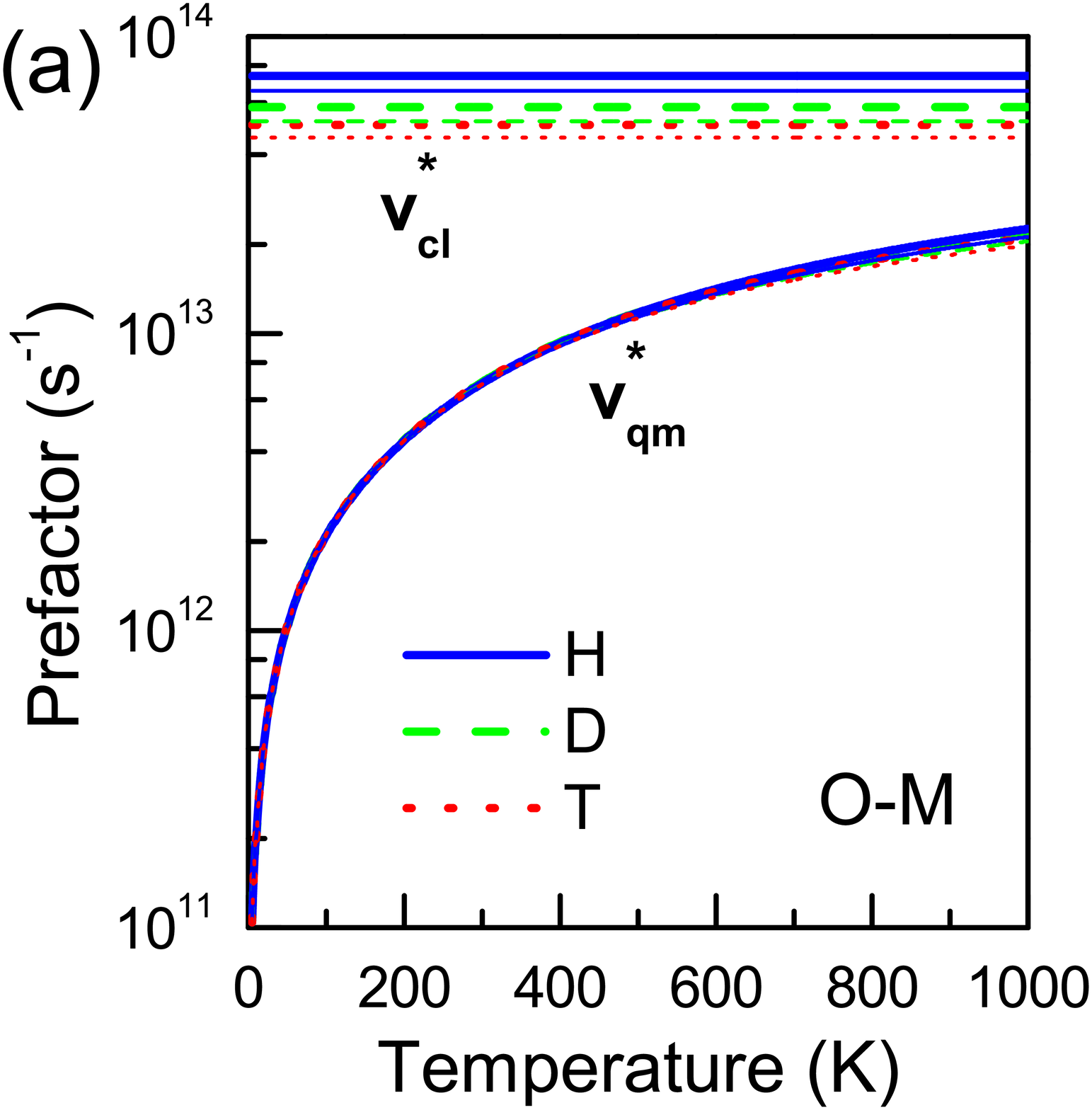}}
\scalebox{0.18}[0.18]{\includegraphics{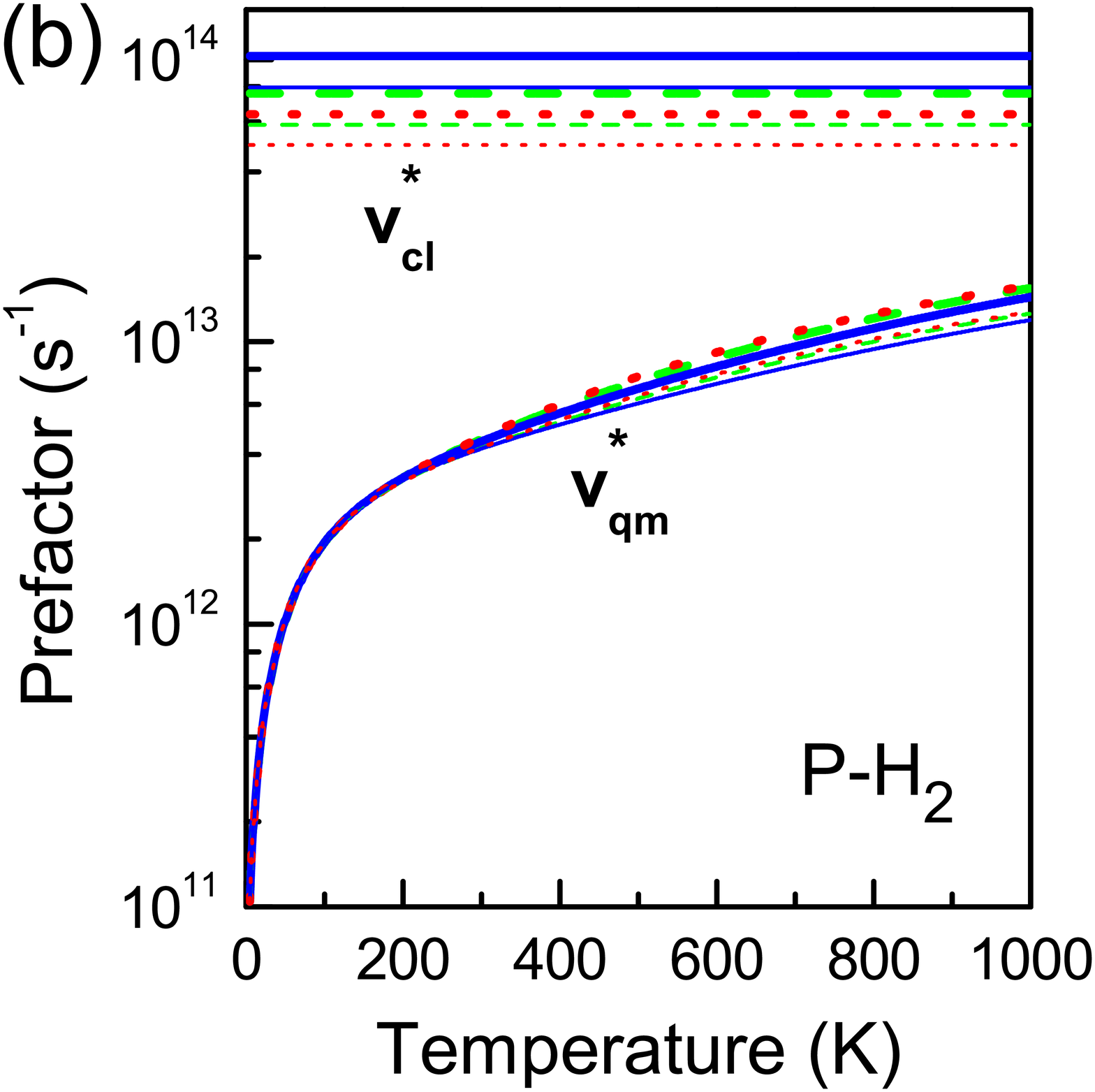}}
\caption{\label{prefactors}(Color online) The variation of the
$v^*_{qm}$ and $v^*_{cl}$ with respect to temperature in the S3 and
S5 cases for (a) O-M and (b) P-H$_2$ transitions of H, D and T
dimers. The prefactors for the S3 case are plotted with thicker
lines, and those for the S5 case with thinner lines.}
\end{figure}

 The small difference between the vibrational properties of
hydrogen dimers in the S3 and S5 cases has also been found by
comparing the prefactors for the O-M and P-H$_2$ transitions of H, D
and T dimers in these two cases (Fig. \ref{prefactors}). In the
plotted temperature range, the $v_{cl}^*$ differences between these
two cases are within 20\%, and the $v_{qm}^*$ differences within
15\%. Since $v$ is just linearly dependent on $v^*_{qm}$ (Equ.
\ref{qm_hTST}), a small difference ($< 15\%$) in the value of
$v_{qm}^*$ will not result in any significant deviation in the
kinetic properties, e. g. the TD spectra. Consequently, the
vibrational frequencies from the S3 case are used in the simulations
of the kinetic properties of hydrogen dimers in all cases, which is
a quite accurate treatment and brings much convenience to the
numerical simulations. Seen from Fig. \ref{prefactors}, $v^*_{cl}$
decreases with increasing the hydrogen mass, while the variation of
$v^*_{qm}$ with the hydrogen mass is different. The $v^*_{qm}$ for
the O-M transition is nearly invariant with the hydrogen mass in the
plotted temperature range, while that for the P-H$_2$ transition
increases with the hydrogen mass. These isotope effects in
$v^*_{cl}$ and $v^*_{qm}$ of the hydrogen dimer are the same as
those for the desorption and diffusion of the hydrogen monomer on
graphene,\cite{lfhuang1} which has been understood from the isotope
effect of the spectra of the localized vibrational modes of hydrogen
and analyzed in detail in Ref. \onlinecite{lfhuang1}. The hydrogen
dimers are kinetically active in the temperature range from 350 to
650 K, above which the dimers have been completely desorbed (shown
below). In this temperature range, the $v^*_{qm}$s for the O-M
transition vary from 8 to $15\times10^{12}$ s$^{-1}$, and those for
the P-H$_2$ transition vary from 5 to $9\times10^{12}$ s$^{-1}$.

\par
\begin{figure}[!h!t!bp] 
\scalebox{0.18}[0.18]{\includegraphics{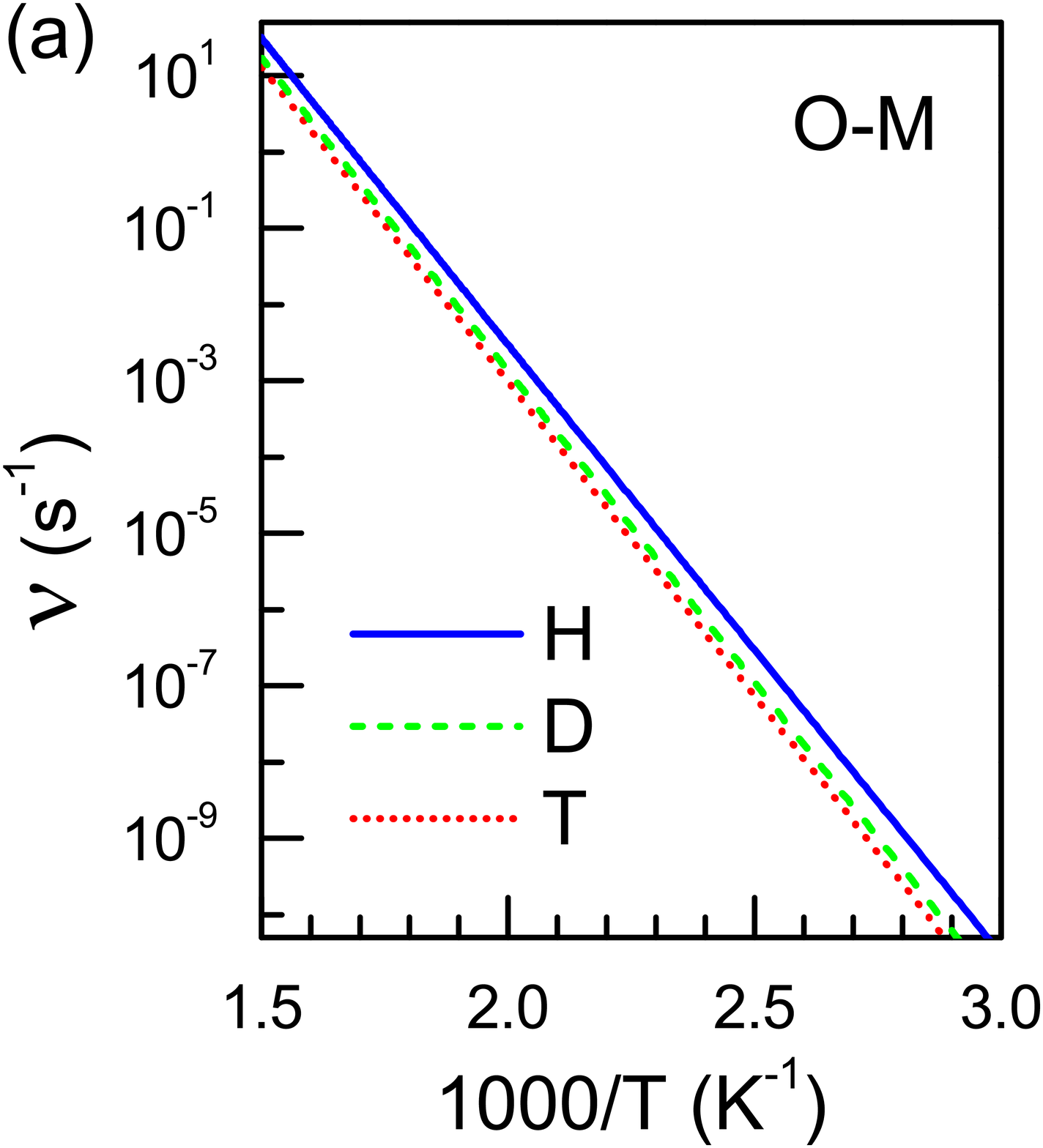}}
\scalebox{0.18}[0.18]{\includegraphics{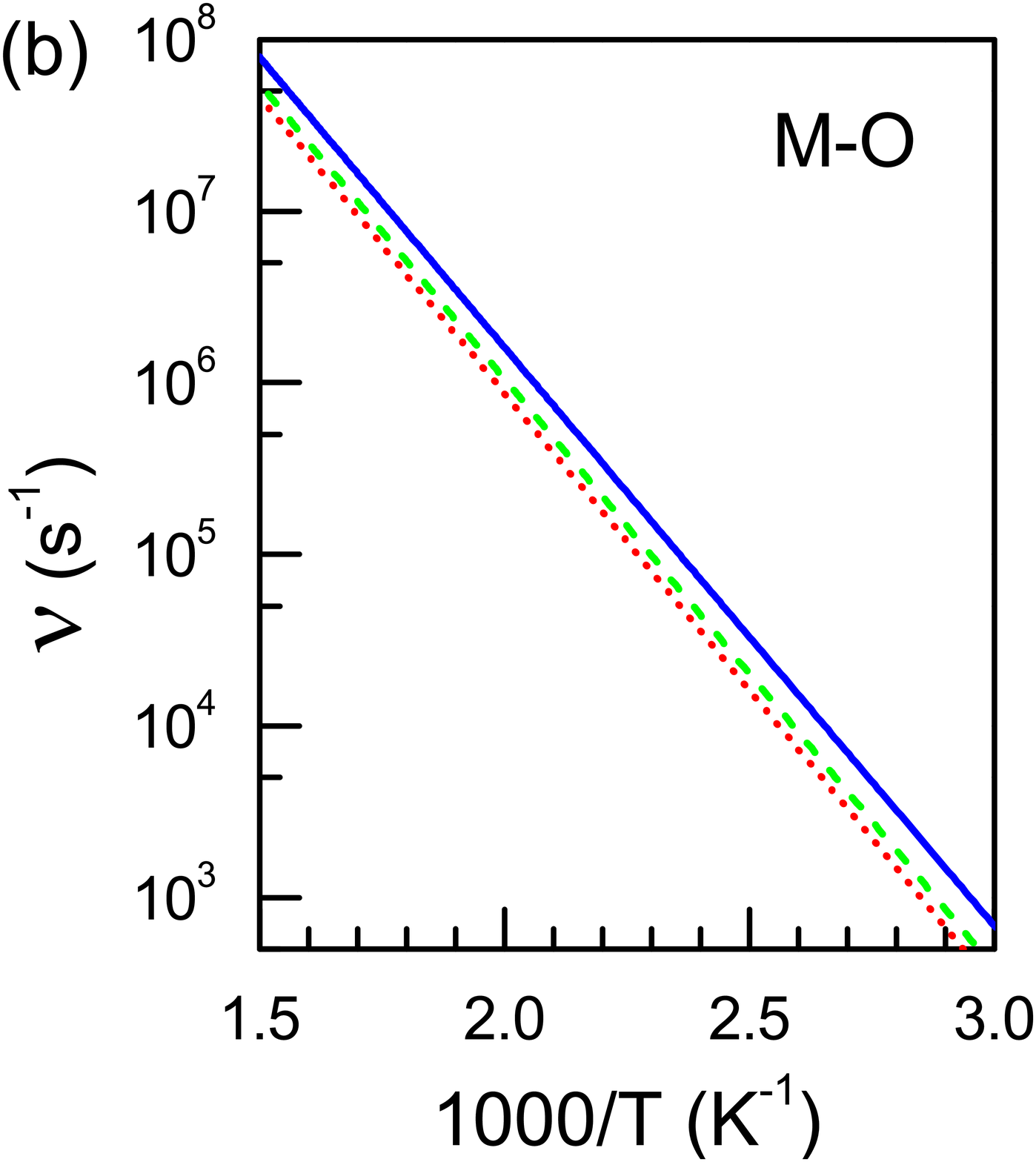}}
\scalebox{0.18}[0.18]{\includegraphics{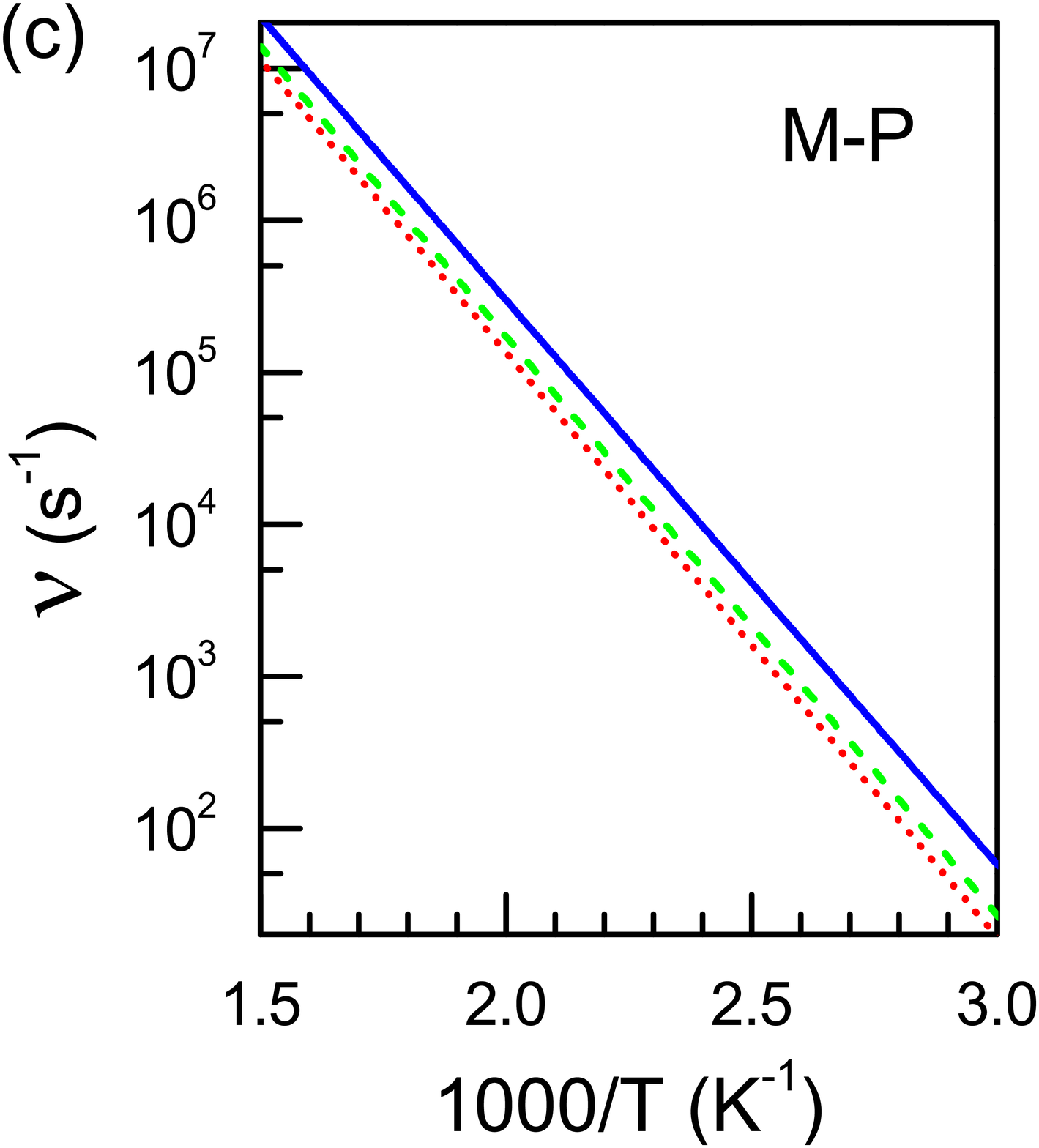}}
\scalebox{0.18}[0.18]{\includegraphics{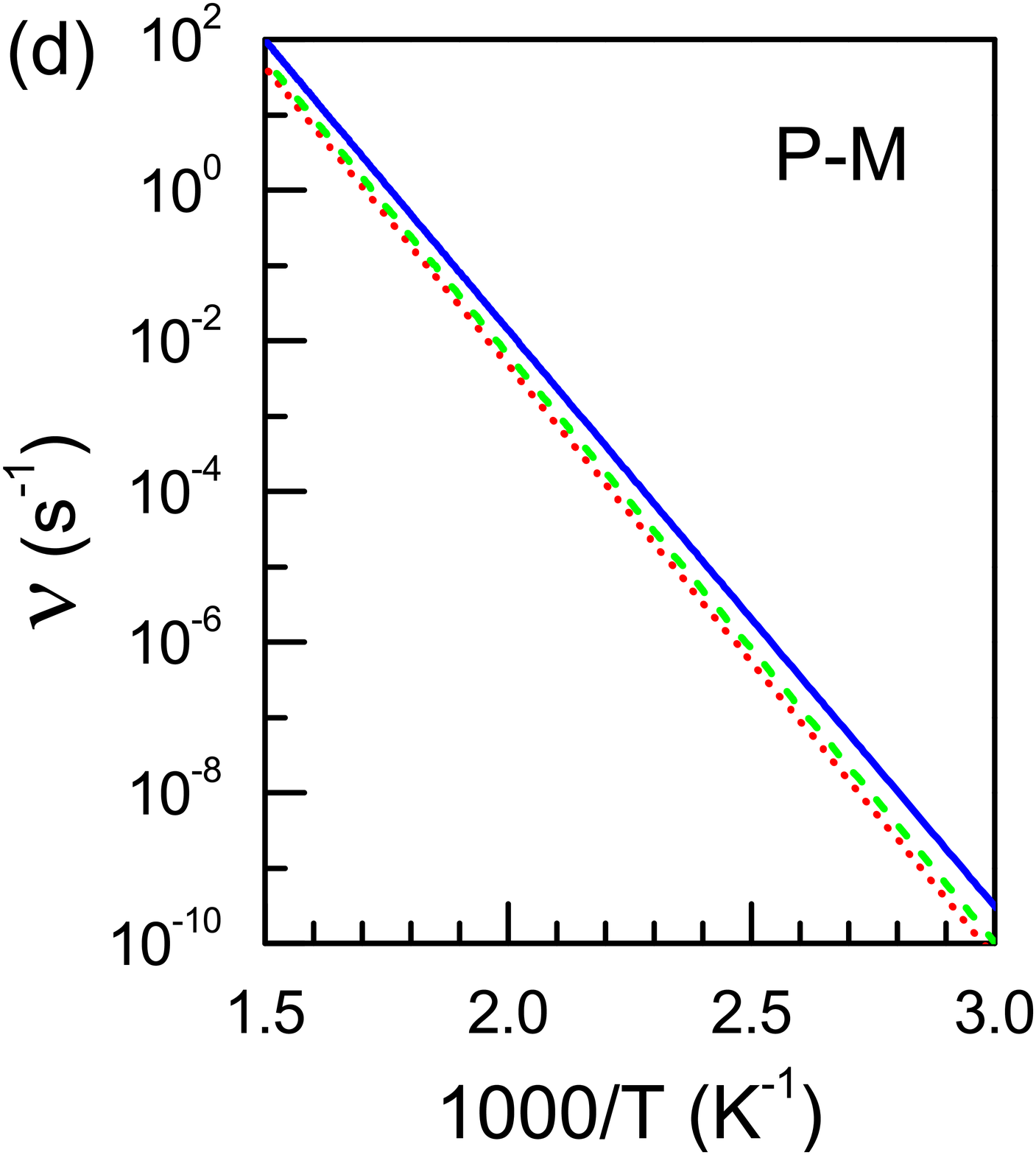}}
\scalebox{0.18}[0.18]{\includegraphics{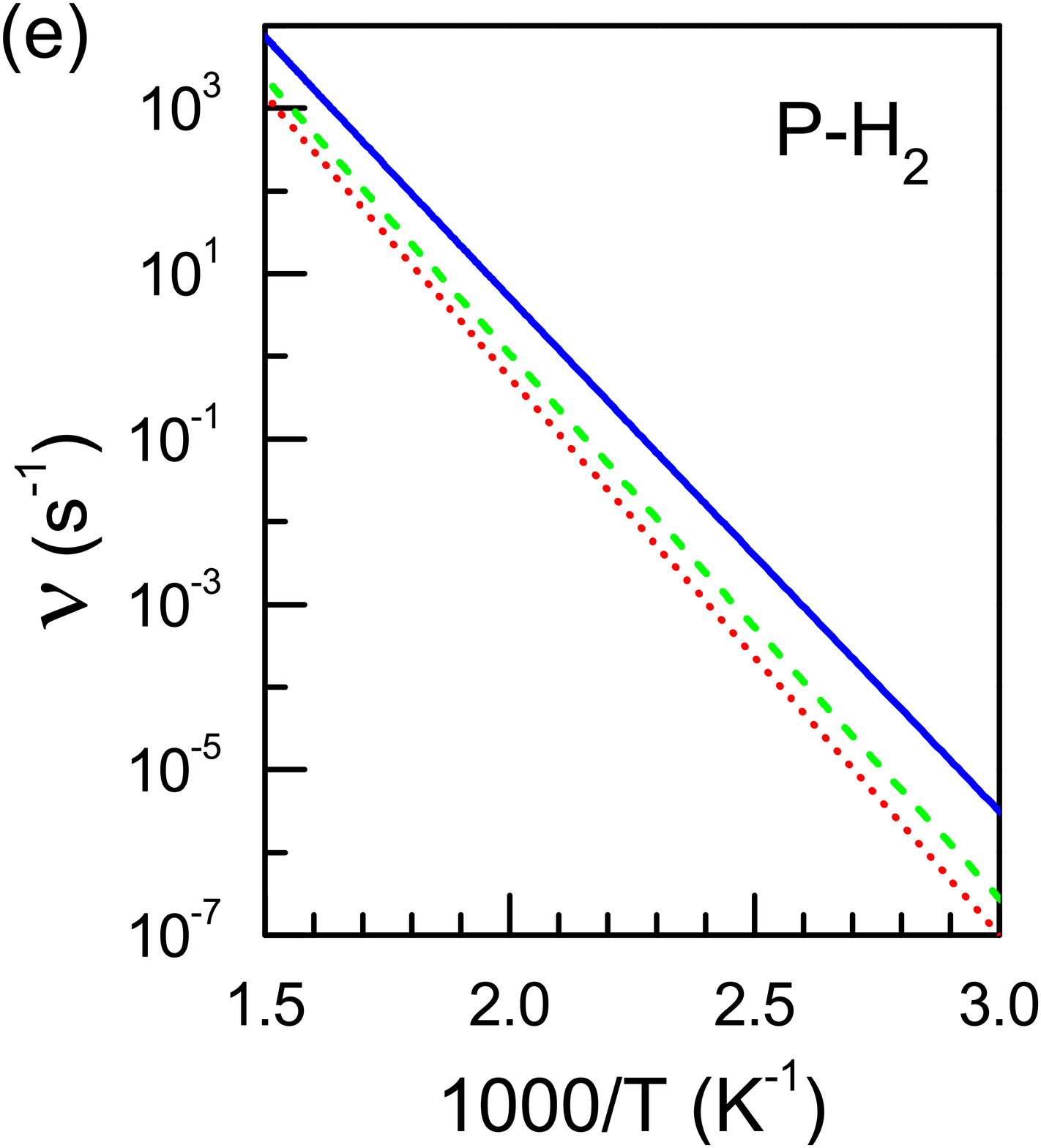}}
\scalebox{0.18}[0.18]{\includegraphics{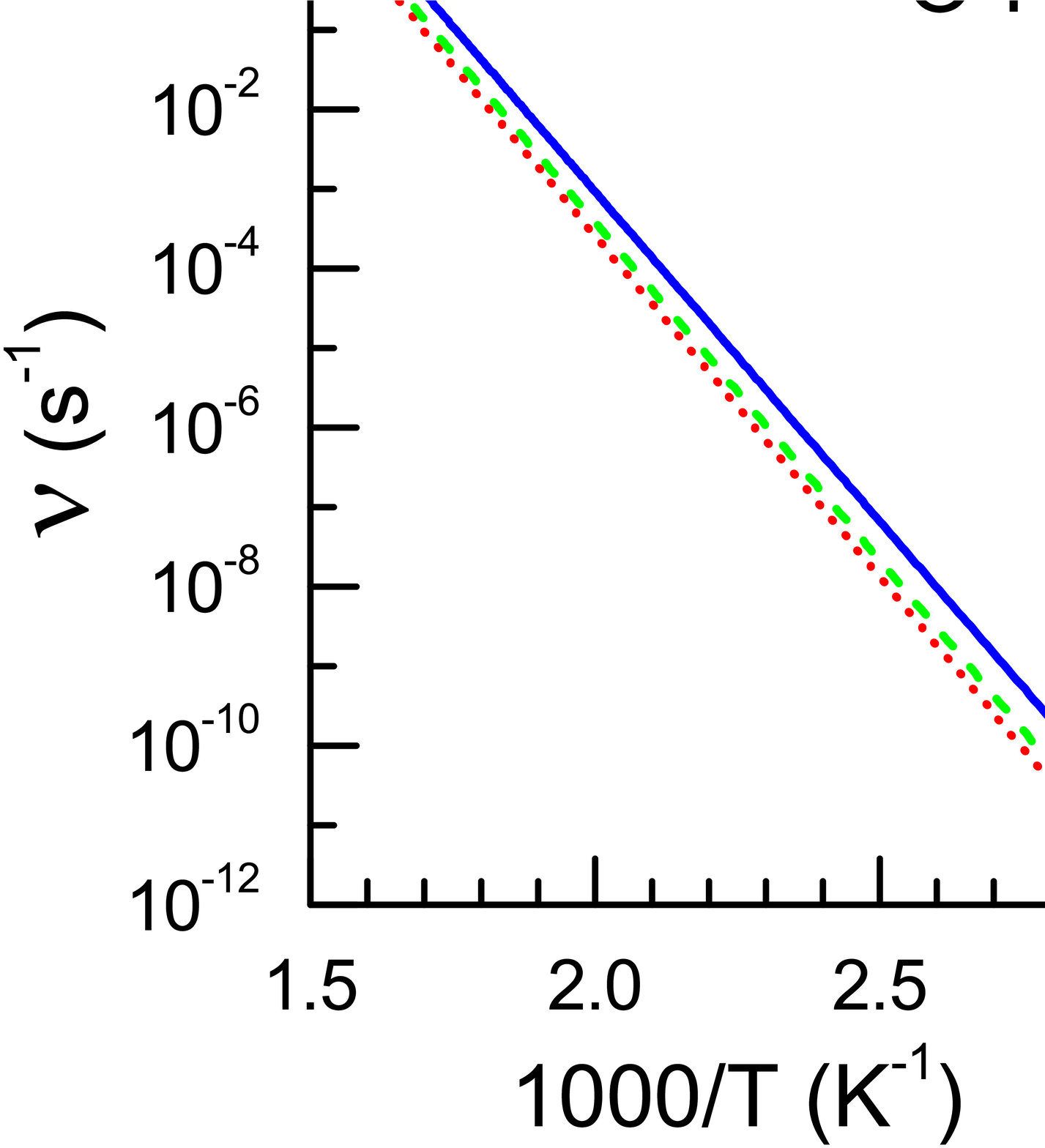}}
\scalebox{0.18}[0.18]{\includegraphics{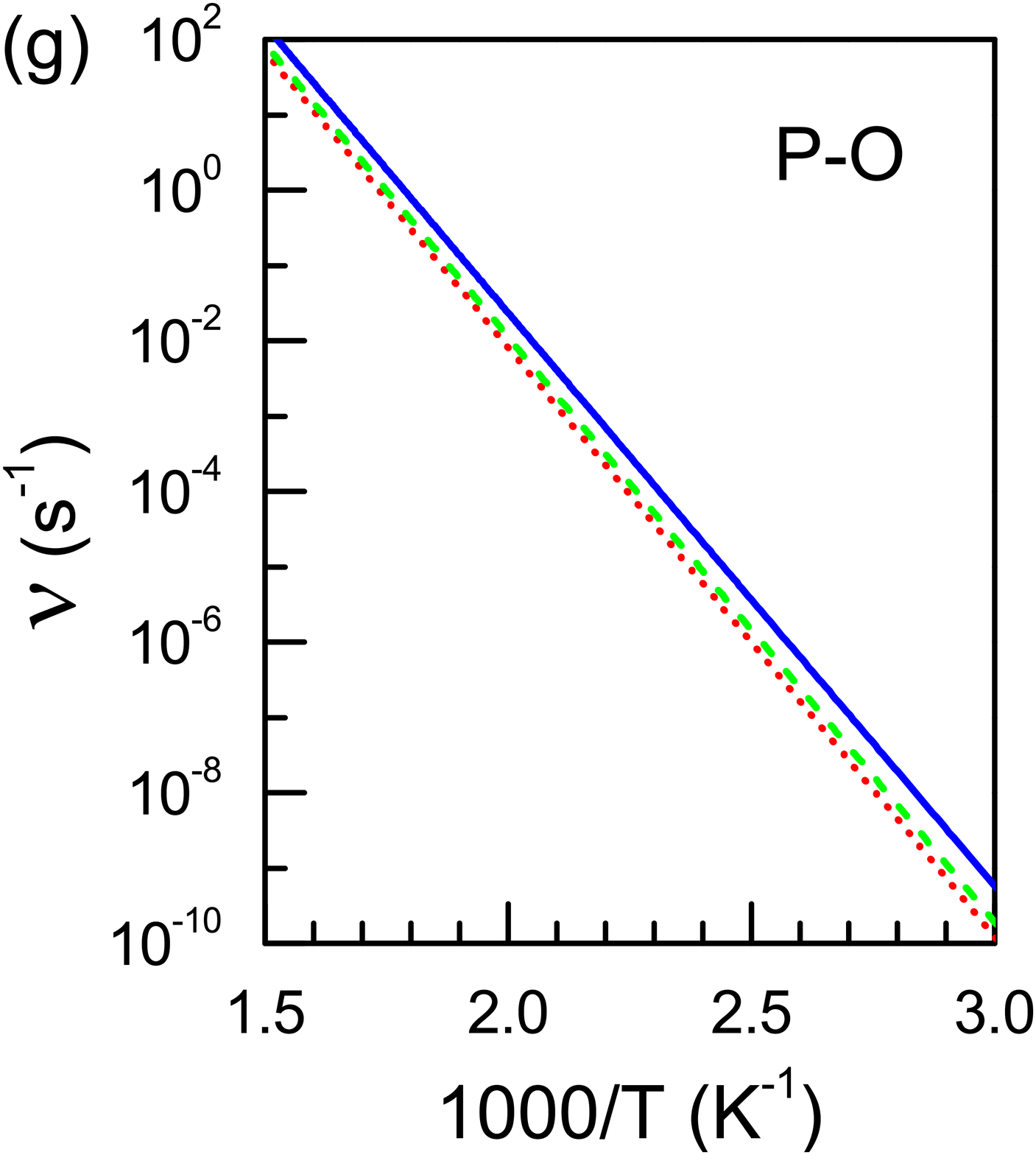}}
\caption{\label{jfreq}(Color online) The variation of $v$ with
respect to the inverse of temperature for the (a) O-M, (b) M-O, (c)
M-P, (d) P-M, (e) P-H$_2$, (f) O-P and (g) P-O transitions of H, D
and T dimers. The corresponding $v_{qm}^*$s are used as the
prefactors.}
\end{figure}

 The calculated jump frequencies for the transitions of H, D
and T dimers on the S6-P supercell are shown in Fig. \ref{jfreq} (a
- e). The jump frequency of a transition in other cases ($v^X$, X =
S3, S4, S5, S6-O) can be deduced from the corresponding $v$ in the
S6-P case by the equation
\begin{eqnarray}
v^X=v^{S6-P}\exp{({-\frac{\delta\epsilon}{k_BT}})}
\end{eqnarray}
where $\delta\epsilon$ is the difference of the ${\Delta}E_p$ for a
transition in the X case and the S6-P case. The values of the
${\Delta}E_p$s for all the transitions in all the cases are listed
in Tab. \ref{potential_barrier}. The isotope effect in $v$ is that
it decreases with increasing the hydrogen mass, which is due to the
isotope effect in ${\Delta}F_{vib}(0)$ (Tab. \ref{ZPE_corrections})
and is the same as that of the hydrogen monomer on
graphene\cite{lfhuang1}. In the temperature range from 350 to 650 K,
the jump frequencies for the M-O and M-P transitions are at least 4
orders larger than those for the O-M, P-M and P-H$_2$ transitions.
Because $v$ is exponentially dependent on $E_{ac}$ or ${\Delta}E_p$
(Equ. \ref{qm_hTST} and \ref{jfreq}) and the ${\Delta}E_p$s for the
M-O and M-P transitions are much less than those for the O-M, P-M
and P-H$_2$ transitions. The life time of the M-dimer
($\tau_M\sim\frac{1}{v_{M-P}+v_{M-O}}$) in the S6-P case is less
than 1.0 s at 300 K, and decreases exponentially with temperature.
And $\tau_M$ is much less in other cases due to their lower
$\Delta{E}_{p}$s for the M-P and M-O transitions (Tab.
\ref{potential_barrier}). Thus, although the M-dimer is an important
configuration in the kinetic transitions of a hydrogen dimer on
graphene, it has not been observed in
experiments\cite{hornekaer96,andree425} and theoretical
simulations.\cite{dumont77} Furthermore, $v_{P-H_2}$ transition is
at lest 2 orders larger than $v_{O-M}$, which indicates that the
O-dimer is kinetically more stable than the P-dimer on graphene.

\subsection{The simulation of the TD spectra}

\par In the optimal path of O-M-P-H$_2$, the jump frequencies for
the M-O ($v_{M-O}$) and M-P ($v_{M-P}$) transitions are at least 4
orders of magnitude larger than those for other transitions in the
temperature range from 350 to 650 K. Thus, it is reasonable to
assume that when a dimer succeeds in jumping from O-dimer state to
M-dimer state, it will immediately turn to be O-dimer or P-dimer
state. Following the treatment by Toyoura in Ref.
\onlinecite{toyoura78} for the diffusion of lithium in the
intercalated graphite, the mean time of the total jump is the sum of
those of the individual jump steps, $\tau = \tau_1 + \tau_2$, where
$\tau_1 = (g_{O-M}v_{O-M})^{-1}$ and $\tau_2 = (g_{M-P}v_{M-P} +
g_{M-O}v_{M-O})^{-1}$, with the $g_{path}$ of each transition path
being accounted. An O-P transition is accomplished if the M-dimer
finally succeeds in jumping to be a P-dimer. Then the jump frequency
for this O-P transition is given by
\begin{eqnarray}{\label{v_O_P_1}}
v_{O-P}&=&\frac{g_{M-P}v_{M-P}}{g_{M-O}v_{M-O}+g_{M-P}v_{M-P}}\times\frac{1}{\tau_1+\tau_2}\\
\label{v_O_P_2} &=&g_{O-P}\frac{v_{M-P}}{v_{M-O}+v_{M-P}}v_{O-M}
\end{eqnarray}
where Equ. \ref{v_O_P_2} is exact only for the isolated dimers on
graphene and $g_{O-P}$ ($=g_{O-M}g_{M-P}/2$) equals 4 there;
$\frac{g_{M-P}v_{M-P}}{g_{M-O}v_{M-O}+g_{M-P}v_{M-P}}$ or
$\frac{v_{M-P}}{v_{M-O}+v_{M-P}}$ is the success probability of the
M-P jump from a M-dimer. $g_{O-P}$ is regarded as the path
degeneracy of the O-P transition, and Equ. \ref{v_O_P_2} is used for
non-isolated dimers on graphene where $g_{O-P}$ is treated as a
variable. Likewise, the jump frequency for the P-O transition is
given by
\begin{eqnarray}{\label{v_P_O}}
v_{P-O} = g_{P-O}\frac{v_{M-O}}{v_{M-O}+v_{M-P}}v_{P-M}
\end{eqnarray}
The interaction between neighboring hydrogen dimers tends to lower
the values of $g_{O-P}$ and $g_{P-O}$ from 4. Both of $g_{O-P}$ and
$g_{P-O}$ are taken to be equal in the simulation of the TD spectra
for convenience. In Fig. \ref{jfreq} (f) and (g), the calculated
$v_{O-P}$s and $v_{P-O}$s for H, D and T dimers are shown, with
$g_{O-P}$ and $g_{P-O}$ both being taken an intermediate value of 2.
$v_{O-P}$ and $v_{P-O}$ are mainly determined by $v_{O-M}$ and
$v_{P-M}$, respectively, with the multiplying factors in Equ.
\ref{v_O_P_2} and \ref{v_P_O} just varying from 0.5 to 1.0 in the
temperature range from 350 to 650 K. $v_{O-P}$ and $v_{P-O}$
decrease with increasing the hydrogen mass, the same isotope effect
as other transition frequencies in the optimal path of O-M-P-H$_2$
(Fig. \ref{jfreq}).

\par The ratio of O-dimer and P-dimer ($n_{O}:n_{P}$) at the starting
moment (t=0) was reported as 0.15\cite{hornekaer96},
0.2\cite{gavardi477} and 0.47\cite{dumont77}, respectively. In this
report, we take this ratio as 0.2 ($n_O=1.0$ and $n_P=5.0$). If the
interaction between hydrogen dimers approximately acts as an average
background effect, the reaction of the hydrogen dimers on graphene
is first-order, which is consistent with experimental
observations.\cite{zecho42,zecho117} The reaction-rate equations for
the TD spectroscopy of hydrogen dimers on graphene can be expressed
as
\begin{eqnarray}{\label{reaction_rate}}
\frac{dn_O(t)}{dt}&=&-v_{O-P}(T)n_O(t)+v_{P-O}(T)n_P(t)\\
\frac{dn_P(t)}{dt}&=&-[v_{P-O}(T)+v_{P-H_2}(T)]n_P(t)+v_{O-P}(T)n_O(t)\\
n_O(t+{\Delta}t)&=&n_O(t)+\frac{dn_O(t)}{dt}{\Delta}t\\
n_P(t+{\Delta}t)&=&n_P(t)+\frac{dn_P(t)}{dt}{\Delta}t
\end{eqnarray}
where ${\Delta}t$ is a time interval. The desorption rate of
hydrogen dimers, or the hydrogen-gas yield per unit time, is defined
as
\begin{eqnarray}{\label{desorption_rate}}
R_{des}(T,t)=v_{P-H_2}(T)n_P(t)
\end{eqnarray}
where T $={\alpha}t$ and the heating rate ($\alpha$) is set as 1.0
K/s in accordance with
experiments.\cite{hornekaer96,zecho42,zecho117}

\par
\begin{figure}[!h!t!bp] 
\scalebox{0.30}[0.30]{\includegraphics{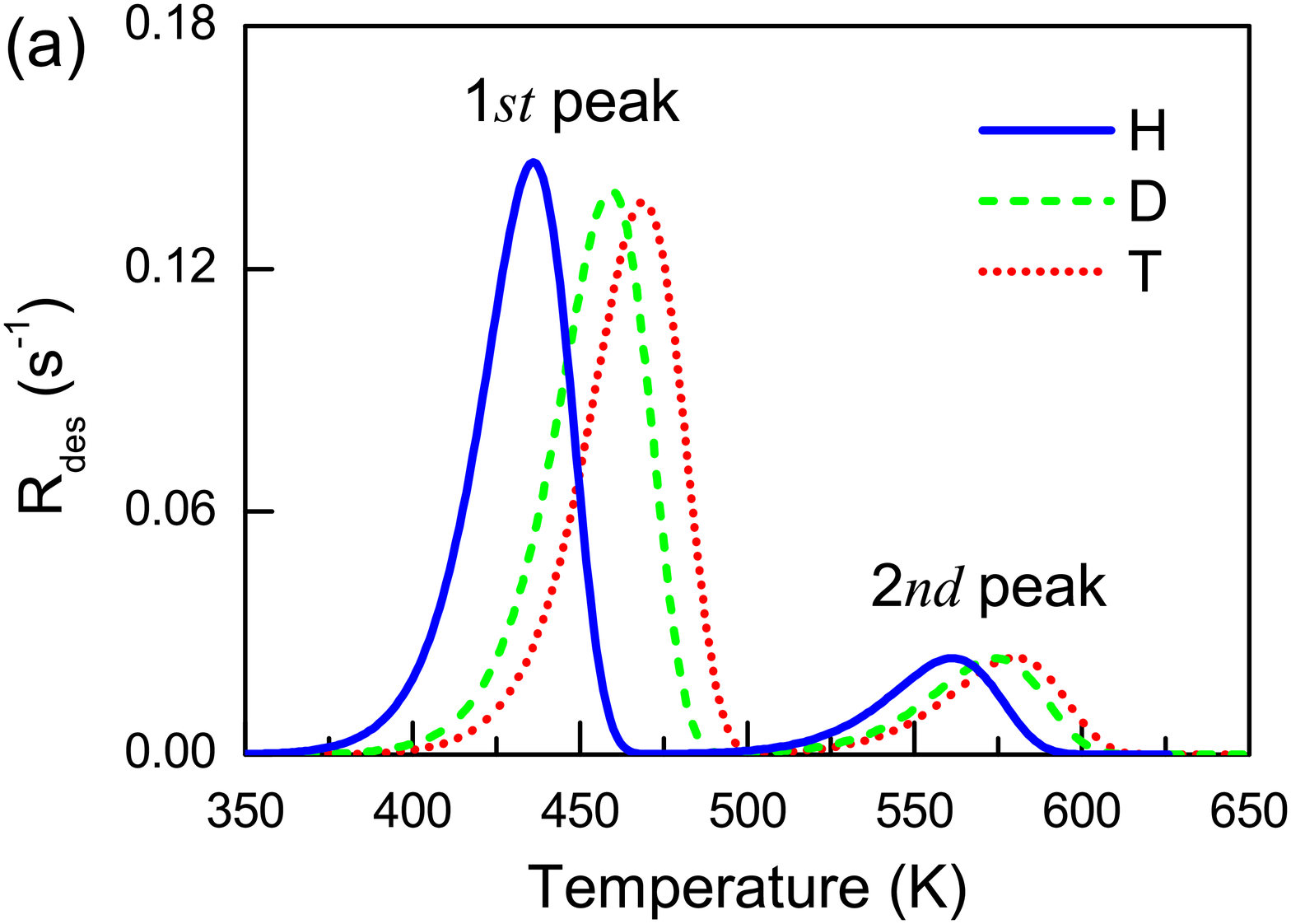}}
\scalebox{0.30}[0.30]{\includegraphics{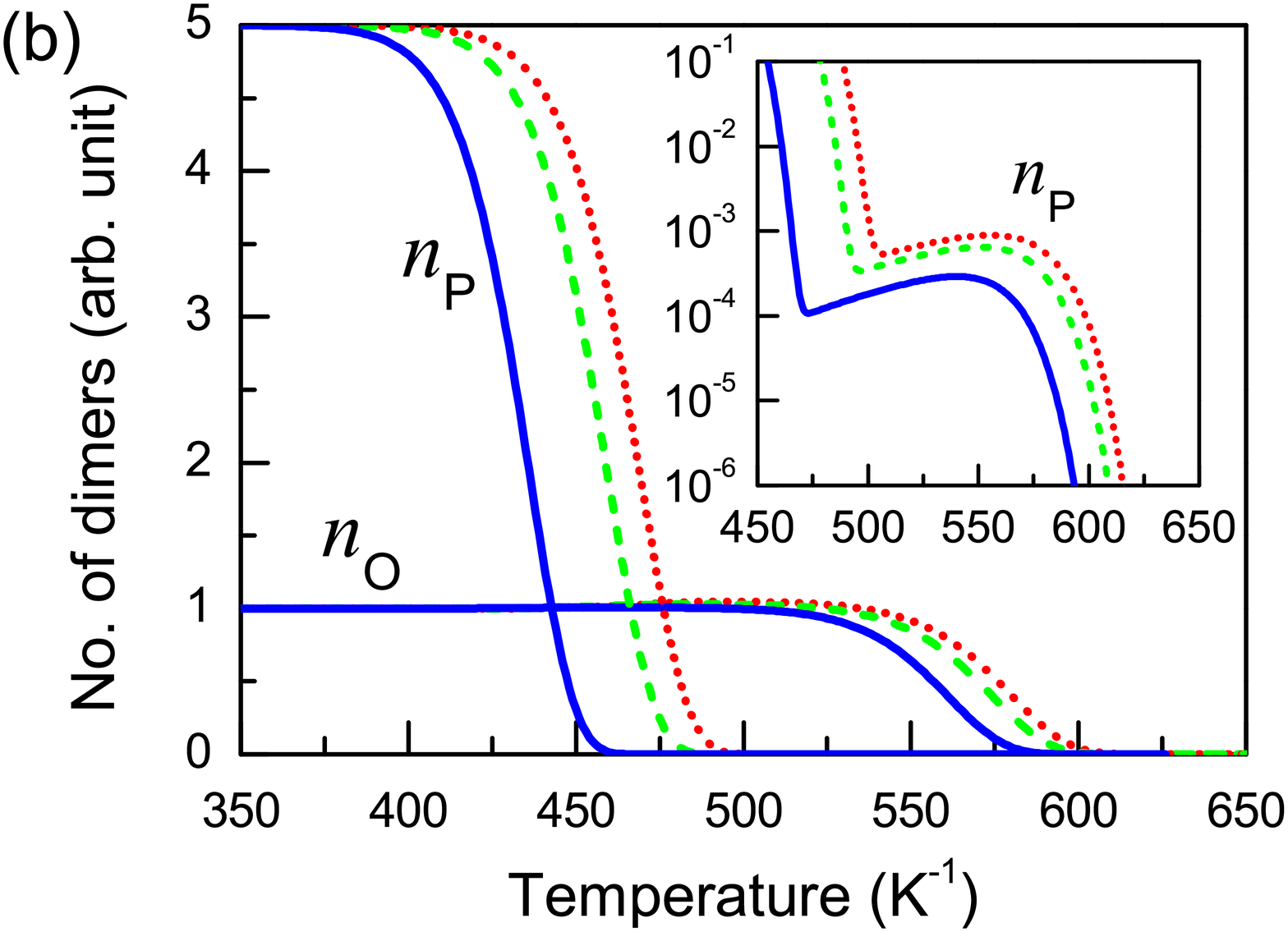}}
\scalebox{0.30}[0.30]{\includegraphics{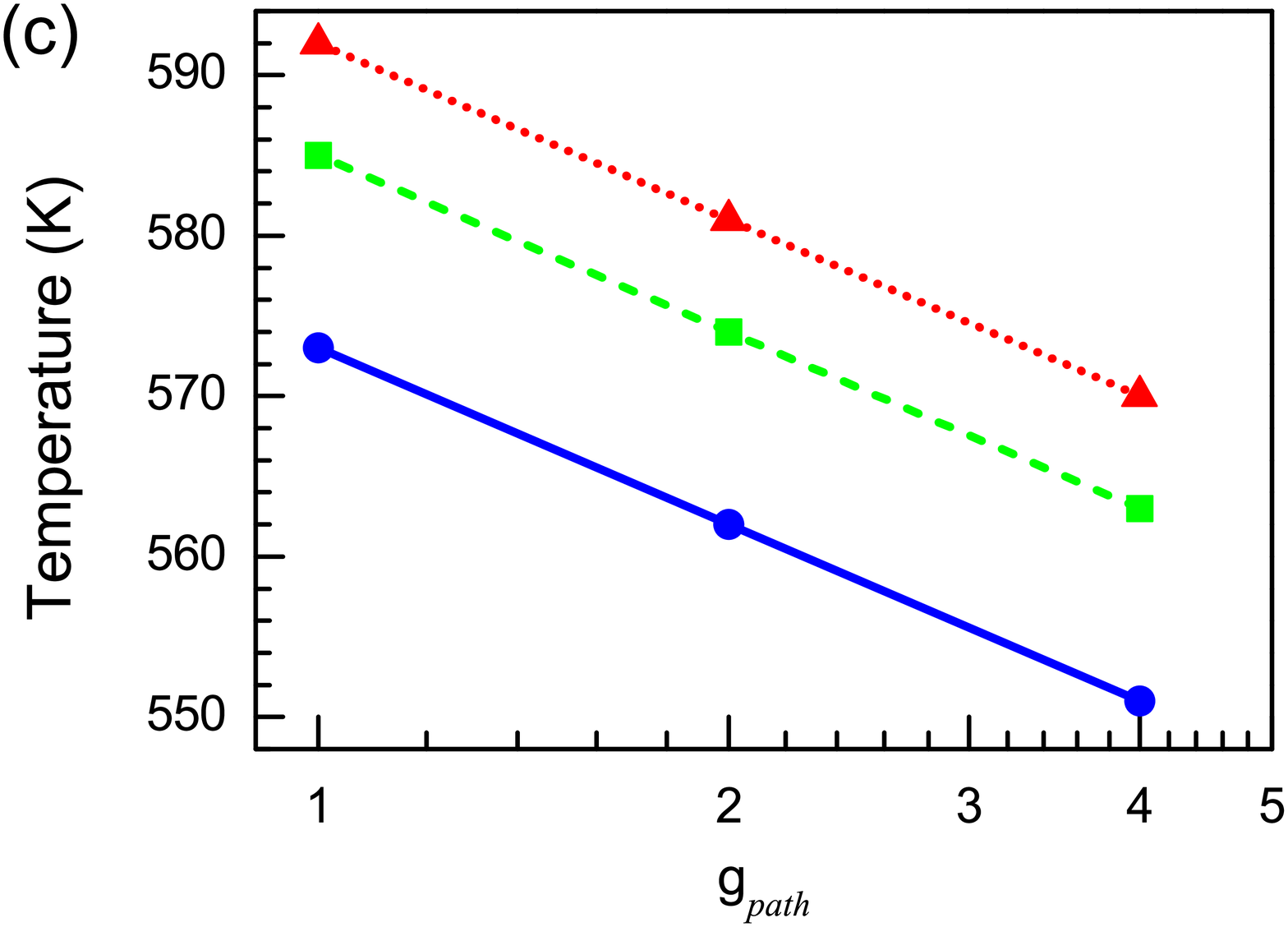}}
\caption{\label{TDS}(Color online) (a) The TD spectra of H, D and T
dimers ($\alpha = 1.0$ K/s) with the starting dimer ratio $n_O(0) :
n_P(0) = 1 : 5$. (b) The variations of $n_P$(T/$\alpha)$ and
$n_O$(T/$\alpha)$. The inset shows the details of the variation of
$n_P$(T/$\alpha$) (in the logarithmic scale) from 450 K to 650 K.
(c) The variations of the positions of the 2{\itshape{nd}}
desorption peak in the TD spectra for H, D and T dimers with respect
to the degeneracy ($g_{path}$) of the diffusing reaction path .
$g_{path}$ is plotted in the logarithmic scale.}
\end{figure}

 The simulated TD spectra for H, D and T dimers in the S6-P case
are shown in Fig. \ref{TDS} (a). The two-peak shaped TD spectra have
been reproduced, with the desorption peaks being at 436 (460, 469)
and 562 (574, 581) K for H (D, T) dimers, respectively. These
simulation results are very close to experimental
measurements.\cite{hornekaer96,zecho42,zecho117} The kinetic
processes behind the TD spectra are described in the following. As
the temperature increases (T$=1.0t$), the P-dimers are first
desorbed through the P-H$_2$ transition above 350 K. At this time,
there are negligible amount of P-dimers transiting to be O-dimers,
because $v_{P-M}$ ($v_{P-O}$) is at lest 2 orders less than
$v_{P-H_2}$. Afterwards, the O-dimers are reduced at higher
temperatures ($> 475$ K) through the O-M-P-H$_2$ transition. The
variations of $n_P(T/\alpha)$ and $n_O(T/\alpha)$ are shown in Fig.
\ref{TDS} (b). When a dimer succeeds in transiting from an O-dimer
through a M-dimer to a P-dimer, it stays as a P-dimer for a short
time of about $1/v_{P-H_2}$ seconds before being desorbed, which is
reflected by the small occupation number of the P-dimers
($n_P(T/\alpha)$) at the temperatures around the 2{\itshape{nd}}
desorption peak (the inset in Fig. \ref{TDS} (b)). The mobilities of
the hydrogen dimer in diffusion and desorption decrease with
increasing the hydrogen mass, which is due to the isotope effect in
$v$ or ${\Delta}F_{vib}(0)$. In experimental
measurements,\cite{hornekaer96,zecho42,zecho117} the positions of
the two desorption peaks in the TD spectra for H (D) dimers are at
445 (490) K and 560 (580) K (the average $\Theta < 0.12$ ML), while
those in the simulated spectra are 436 (460) K and 562 (574) K (the
effective $\Theta = 0.056$ ML). The simulated isotopic differences
of the desorption peaks of H and D dimers are 24 and 12 K for the
1{\itshape{st}} and 2{\itshape{nd}} peaks, respectively, in
comparison with the experimental measurements of 45 and 20 K,
respectively. These little deviation between our theoretical
simulations and experimental results maybe come from that a small
amount of uncertainty has been introduced in experiments when
estimating the desorption-peak positions from the somewhat chaotic
TD spectra of H,\cite{zecho117} or from the isotope effect in the
effective $\Theta$ of the hydrogen-covered spots on graphene in
experiments. The latter possibility is understandable since that the
thermodynamic stability of the hydrogen dimer on graphene should be
hydrogen-mass dependent, thus, the distribution of hydrogen dimers
on graphene should be hydrogen-mass dependent. This is like the
isotope effect of $\Theta_{sat}$, which are about 0.3 and 0.4 ML for
H and D on graphene, respectively.\cite{zecho117} The difference in
the effective $\Theta$ for H and D dimers results in the difference
in the interaction between dimers and then the difference in
${\Delta}E_p$. Based on this point, the isotope effect in the
experimental TD spectra of hydrogen dimers consists of the
contributions from the isotope effects in $v$ and $\Theta$. The
simulations here, however, just consider the isotope effect in $v$.

\par In order to observe the effects of $\Theta$ and the structural configuration on the
TD spectra, the ${\Delta}E_p$s for the transitions of O-M, M-O, M-P,
P-M and P-H$_2$ in all the cases considered here (Tab.
\ref{potential_barrier}) are used to simulate the TD spectra. The
positions of the two desorption peaks in each case are listed in
Tab. \ref{T_peaks}. The 1{\itshape{st}} desorption peak shifts
nearly linearly with the ${\Delta}E_p$ of the P-H$_2$ transition by
34 K/0.1 eV, while the 2{\itshape{nd}} peak does not have a simple
monotonic variation due to the complex transitions of hydrogen dimer
behind but still generally increases with $\Delta{E}_p$.
Additionally, the interaction between dimers also tends to lower the
degeneracy of the diffusing reaction paths, $g_{O-P}$ and $g_{P-O}$,
which are taken equally here and together denoted as $g_{path}$
below. The position of the 1{\itshape{st}} desorption peak is
obviously not dependent on the diffusing $g_{path}$. Seen from Fig.
\ref{TDS} (c) whereas, the positions of the 2{\itshape{nd}}
desorption peaks for H, D and T dimers are nearly logarithmically
dependent on $g_{path}$, which is approximately expressed as
\begin{eqnarray}{\label{T2_vs_log_g}}
T(g_{path})=T(1)-36\times\lg{(g_{path})}
\end{eqnarray}
where T(1) is the peak position when $g_{path}=1$. And T($g_{path}$)
increases by about 11 K with $g_{path}$ decreasing by a half.

\par From the analysis above, the interaction between hydrogen dimers
on graphene is coordinated with  the effective hydrogen coverage and
the structural configuration. Such interaction affects the potential
barriers and the reaction-path degeneracies of the hydrogen-dimer
transitions on graphene. As a result, the jump frequencies and TD
spectra vary with the hydrogen coverage and the structural
configuration. However, the localized vibrational properties of
hydrogen dimers are not significantly influenced by the hydrogen
coverage.

\section{CONCLUSIONS}
\par The thermodynamic and kinetic properties of H, D and T dimers on graphene have been simulated
with a composite ab initio method consisting of density functional
theory, density functional perturbation theory, harmonic transition
state theory, and reaction rate theory. The effects of the hydrogen
coverage and the structural configuration on the thermodynamic and
kinetic properties have been revealed by varying the coverage from
0.040 to 0.111 ML (1.0 ML $= 3.8\times10^{15}$cm$^{-2}$). It has
been found that the interaction between hydrogen dimers influences
the effective hydrogen coverage and the configuration of hydrogen
dimers deposited on graphene and affect the thermodynamic and
kinetic properties ($E_{ads}$, ${\Delta}E_p$, $v(T)$ and
$R_{des}(T,t)$) of those dimers. The vibrational zero-point energy
correction (${\Delta}F_{vib}$(0)) and the jump frequency ($v$) both
decrease with increasing the hydrogen mass. However, the isotope
effect in the positions of the desorption peaks in the TD spectra is
inverted. In a word, the mobility of the hydrogen dimer in
desorption and diffusion decreases with increasing the hydrogen
mass. The simulated TD spectra (heating rate $\alpha = 1.0$ K/s) are
quite close to experimental measurements, and the effect of the
dimer interaction has been clarified in detail via the potential
barrier ${\Delta}E_p$ in various cases and the reaction-path
degeneracy $g_{path}$.

\begin{acknowledgments}
The first author (Huang) wish to thank Liv Horkek{\ae}r for helpful
email exchanges. This work was supported by the special Funds for
Major State Basic Research Project of China (973) under grant No.
2007CB925004, Knowledge Innovation Program of Chinese Academy of
Sciences under grant No. KJCX2-YW-N35, and Director Grants of
CASHIPS. Part of the calculations were performed in Center of
Computational Science of CASHIPS and the Shanghai Supercomputer
Center.
\end{acknowledgments}

\bibliography{basename of .bib file}

\begin{table}[!h!t!bp] 
\caption{\label{E_ads}The hydrogen coverage ($\Theta$) and the
adsorption energy of the para-dimer ($E_{ads}^{P}$) and ortho-dimer
($E_{ads}^{O}$) in various cases (1.0 ML$ = 3.8\times10^{15}$
cm$^{-2}$).}
\begin{ruledtabular}
\begin{tabular}{cccc}
Supercell&$\Theta$ (ML)&$E_{ads}^{P}$ (eV)&$E_{ads}^{O}$
(eV)\\\hline
S3&0.111&2.714&2.555\\
S4&0.063&2.562&2.564\\
S5&0.040&2.538&2.518\\
S6-P&0.056&2.604&2.758\\
S6-O&0.056&2.578&2.652\\
\end{tabular}
\end{ruledtabular}
\end{table}

\begin{table}[!h!t!bp] 
\caption{\label{potential_barrier}The reaction-path degeneracy
($g_{path}$) for an isolated hydrogen dimer on graphene, the
potential barrier (${\Delta}E_p$) for the transitions of a hydrogen
dimer on various graphene supercells.}
\begin{ruledtabular}
\begin{tabular}{ccccccc}
& &\multicolumn{5}{c}{$\Delta{E}_p$ (eV)}\\\cline{3-7}\\
Path&$g_{path}$&S3&S4&S5&S6-P&S6-O\\\hline
O-M&4&1.679&1.699&1.664&1.690&1.713\\
M-O&2&0.405&0.510&0.596&0.689&0.609\\
M-P&2&0.351&0.538&0.607&0.773&0.660\\
P-M&4&1.792&1.711&1.674&1.620&1.690\\
P-A&2&1.821&1.637&1.602&1.605&1.636\\
A-P&1&0.223&0.449&0.525&0.558&0.587\\
M-B&2&0.605&0.806&0.836&0.962&0.912\\
B-M&2&1.106&1.139&1.110&1.121&1.113\\
O-H$_2$&1&2.280&2.228&2.360&2.214&2.262\\
H$_2$-O&-&4.414&4.298&4.470&4.191&4.345\\
M-H$_2$&1&0.801&0.934&0.994&1.149&1.039\\
H$_2$-M&-&4.191&4.187&4.143&4.126&4.227\\
P-H$_2$&1&1.456&1.358&1.336&1.461&1.414\\
H$_2$-P&-&3.398&3.448&3.461&3.593&3.573\\
\end{tabular}
\end{ruledtabular}
\end{table}

\begin{table}[!h!t!bp] 
\caption{\label{ZPE_corrections}The vibrational zero-point energy
correction (${\Delta}F_{vib}(0)$, in eV) for the transitions of H, D
and T dimers in the optimal path (O-M-P-H$_2$) in the cases of S3
and S5.}
\begin{ruledtabular}
\begin{tabular}{ccccccc}
&\multicolumn{3}{c}{S3}&\multicolumn{3}{c}{S5}\\\cline{2-4}\cline{5-7}\\
Path&H&D&T&H&D&T\\\hline
O-M&0.141&0.107&0.093&0.135&0.102&0.088\\
M-O&0.038&0.028&0.017&0.051&0.034&0.027\\
M-P&0.053&0.032&0.023&0.079&0.053&0.042\\
P-M&0.138&0.106&0.093&0.144&0.111&0.075\\
P-H$_2$&0.257&0.187&0.157&0.251&0.184&0.156\\
\end{tabular}
\end{ruledtabular}
\end{table}

\begin{table}[!h!t!bp] 
\caption{\label{T_peaks}The positions (in K) of the two desorption
peaks in the TD spectra of H, D and T dimers in various cases
($\alpha = 1.0 $ K/s).}
\begin{ruledtabular}
\begin{tabular}{ccccccc}
&\multicolumn{3}{c}{1st desorption peak}&\multicolumn{3}{c}{2nd desorption peak}\\\cline{2-4}\cline{5-7}\\
Supercell&H&D&T&H&D&T\\\hline
S3&434&458&468&534&536&551\\
S4&401&424&434&551&563&569\\
S5&393&417&427&537&548&554\\
S6-P&436&460&469&562&574&581\\
S6-O&420&443&453&562&573&579\\
\end{tabular}
\end{ruledtabular}
\end{table}

\end{document}